\documentclass[reprint,amsmath,amssymb,aps]{revtex4-1}

\usepackage{graphicx}
\usepackage{dcolumn}
\usepackage{bm}
\usepackage{subfigure}

\usepackage{xcolor} 

\begin{document}

\preprint{APS/123-QED}

\title{Community structure revealed by phase locking}

\author{Ming-Yang Zhou$^1$}

\author{Zhao Zhuo$^1$}
\email{zhzh7532@mail.ustc.edu.cn}

\author{Shi-min Cai$^2$}

\author{Zhongqian Fu$^1$}%

\affiliation{$^{1}$Department of Electronic Science and Technology, University of Science and Technology of China, Hefei 230027, China.\\
 $^{2}$Web Sciences Center, School of Computer Science and Engineering, University of Electronic Science and Technology of China, Chengdu 611731, P. R. China}

\date{\today}

\begin{abstract}
Community structure can naturally emerge in paths to synchronization, and scratching it from the paths is a tough issue that accounts for the diverse dynamics of synchronization. In this paper, with assumption that the synchronization on complex networks is made up of local and collective processes, we proposed a scheme to lock the local synchronization (phase locking) at a stable state meanwhile suppress the collective synchronization based on Kuramoto model. Through this scheme, the network dynamics only contains the local synchronization, which suggests that the nodes in the same community synchronize together and these synchronization clusters well reveal the community structure of network. Furthermore, by analyzing the paths to synchronization, the relations or overlaps among different communities are also obtained. Thus, the community detection based on the scheme is performed on five real networks and the observed community structures are much more apparent than modularity-based fast algorithm. Our results not only provide a deep insight to understand the synchronization dynamics on complex network but also enlarge the research scope of community detection.
\end{abstract}

\pacs{05.45.Xt, 89.75.Hc, 89.75.Kd}

\maketitle

\textbf{Communities represent the elementary units or function of a system. Detecting effective communities is a key
issue when exploring a complex system. Many previous works based on topology structure and optimized an objective
function to identify communities. The detected communities can't conform to reality fully and many nodes were wrongly classified, which dues to the noise of topology structure. Unlike previous studies, this paper works based
on the dynamics of networked systems and applies synchronization to community detection. A hypothesis is proposed that synchronization consists of two parts: local and collective synchronization. A novel approach is put forward to lock local synchronization and
suppress collective synchronization at the same time in phase synchronization. It obtains amazing results that nodes
in the same community synchronize together, while nodes across different communities have different phases. We therefore
identify communities according to nodes' phases. It achieves good performance on practical networks and the identified communities correspond to reality better than previous methods, which is meaningful
for analyzing the unit function of a system.
}
\begin{center}
\rule[0pt]{2.4in}{0.5pt}
\end{center}
\section{Introduction}

Network science is probably the most attractive field across many research areas. Plenty of complex systems can be represented as networks whose nodes indicate the elements of systems and
edges describe the interaction between the elements \cite{santofortunato2010,porter2009}.
In many networks, nodes with similar property assemble into functional groups. Connections in groups are dense and those groups with high edge density are defined as communities \cite{santofortunato2010}. Thus, identifying communities equals to detecting those dense node clusters. As networks' characteristics and functions have much relation with the topology, detecting communities has become a key challenge in network analysis \cite{santofortunato2010, cocia2011}.Scientists from physics, computer science, applied mathematics, biology and sociology have used various tools and techniques in different research areas to settle the problem \cite{cocia2011, honglei2012, wangbiao2013, zhoum2012}.

Many previous works about community detection focus on the static topology investigation that includes methods based on modularity, K-Clique, betweeness, link clustering and so on \cite{newman2004,palla2005,ahn2010,cocia2011}.
These algorithms perform mainly by optimizing an objective function such as modularity \cite{newman2006} and conductance \cite{lec2009}. However, there is a problem that partitions got by optimizing the objective function don't always reflect the most reasonable community division, \emph{e.g.}, the small communities are often absorbed in large ones in the algorithm based on optimization of modularity \cite{good2010,kehagias2012} due to the confusion or randomness of network topology \cite{albert2012}. Other works apply dynamic processes to revealing community structures since the dynamics and topology are closely related \cite{arenas2008,kim2010,wu2012,arenas2006, toni2011}. For instances, Arenas \emph{et al.} investigated synchronization and found that community structure emerges in the synchronization paths\cite{arenas2008,arenas2006}; G\'{o}mez-Garde\~{n}es \emph{et al.} explored different paths to synchronization  for various community structure, which suggested that synchronization paths could be used to detect communities \cite{jesus2007}; Kim \emph{et al.} found the cluster evolving patterns in the synchronization paths \cite{kim2010}; Wu \emph{et al.} detected communities by adding a repulsing factor to synchronization model \cite{wu2012}; B\"{o}hm \emph{et al.} applied synchronization to data clustering according to the attributes of nodes\cite{bohm2010}; Granell \emph{et al.} added self-loop to each node and identified clusters at different resolution levels \cite{Granell2011, Gomez2008, Granelltools}. Nevertheless, problems still exist in this field due to the confusing hierarchical organization and varying edge density for different clusters.
For example, divisions of most algorithms have several large communities and many extremely small clusters. The small number of large communities almost span the whole network\cite{lec2009}. But practical clusters have more balanced size. So how to apply
synchronization to detect reasonable and even communities in real networks still remains a challenge.

In this paper, we aim to reveal community structures via phase-locked synchronization of Kuramoto model.
The inspiration of phase-locked synchronization relies on the observation that the
synchronization is made up of local and collective dynamics.
The local dynamics reflects substructures in network while the collective dynamics leads to an collective state. Thus, we suppress the collective dynamics and
lock the phase of oscillators into the local dynamics via the synchronizing comparison of the
original network and its corresponding first-order null model network (A null model network follows the same degree distribution with real network, but its edges are fully randomized). The scheme locks local synchronization and suppresses collective synchronization simultaneously. Not only the communities emerge in paths to synchronization, but also the community overlapping phenomenon can also be observed in the synchronizing process.


\section{Phase locking synchronization}
The classic Kuramoto model is adopted to analyze synchronization, of which each node's state is represented
by its phase \cite{arenas2008}.
For a network consisting of $N$ nodes, the evolution of node's phase is determined by its
intrinsic oscillators and the influence of their neighbors,
\begin{equation}\label{kuramoto}
    \frac{d{{\theta }_{i}}}{dt}={{\omega }_{i}}+\frac{c}{N}\sum\limits_{j\in {{\Lambda }_{i}}}{\sin ({{\theta }_{j}}-{{\theta }_{i}})},
\end{equation}
where $\theta_i$, $\omega_i$, $\Lambda_i$ and $c$ are node $i$'s phase, intrinsic frequency,
the set of neighbors and the coupling strength, respectively. If $c$ exceeds a threshold
($C_{threshold}$),
the network can arrive at collective synchronization, otherwise remains in disorder.

In Ref. \cite{arenas2006}, it's reported that the Kuramoto models in network with community structure
firstly achieves cluster synchronization representing communities and then completely synchronizes in time scales.
A hypothesis is first given that synchronization consists of local and collective synchronization mechanisms.
The local synchronization mechanism tends to reveal the community structures while the collective synchronization
mechanisms leads network to a completely synchronized state. The network with community structure
reaches local and collective synchronization simultaneously, while the corresponding null model of
network only arrives at
collective synchronization. Inspired by the original definition of modularity which compares the edge density
differences of original network and the corresponding null model network \cite{newman2006}, we proposal a scheme to
suppress collective synchronization by comparing the synchronization coupling strength of
original network with that of null model network, which reads as
\begin{equation}\label{mykuramoto}
\frac{d{{\theta }_{i}}}{dt}=\sum\limits_{j\in N}{({{a}_{ij}}-{{p}_{ij}})\sin ({{\theta }_{j}}-{{\theta }_{i}})},
\end{equation}
where $a_{ij}$ is the elements of adjacent matrix $A_{N\times N}$ of original network and
$p_{ij}={{k}_{i}}{{k}_{j}}/(2M)$ is the probability of node $i$ and $j$ having a link in the corresponding null model network.
In Eq. \ref{mykuramoto}, $p_{ij}$ plays the role for suppressing collective synchronization. Note that $\omega_i$ and $\frac{c}{N}$ are reduced because they don't affect the evolution of phase in Eq. \ref{mykuramoto}, and $p_{ij}$ may exists even when $a_{ij}=0$. Therefore, considering that each node may interact with all the other ones, we enlarge
the neighbor region into whole network (denoting as $a_{ij}-p_{ij}$ for all pairs of elements).

The modified phase synchronization in Eq.\ref{mykuramoto} affects local synchronization little and suppresses collective synchronization dramatically. When the network
reaches stable state, phase differences of each pair nodes keep unchanged and nodes in the same cluster have similar
phases, suggesting that phases of nodes are locked (i.e., only reaching local synchronization).
In contrary, fully random networks having no community structure will stay in disorder since the
collective synchronization is suppressed. Therefore, community
structures are the foundation of the phase locking state. In other words, the phenomenon occurs only in networks with community structure and no false community will be detected by phase locking in network containing no community structure.

To estimate the performance of synchronization, an order parameter $R$ is introduced \cite{arenas2008} as
\begin{equation}\label{parameter}
R= \left| \frac{1}{N}\sum\limits_{j=1}^{N}{{{e}^{i{{\theta }_{j}}}}} \right|\textcolor{red}{,}
\end{equation}
which evaluates the collective synchronization. Besides $R$, a local order parameter $R_{local}$ \cite{jesus2007} evaluates local synchronization, defined as follows
\begin{equation}\label{localparameter}
{{R}_{local}}=\frac{1}{{M}}\sum\limits_{i}^{{}}{\sum\limits_{j\in {{\Lambda }_{i}}}^{{}}{\cos ({{\theta }_{i}}-{{\theta }_{j}})}},
\end{equation}
where $\Lambda_{i}$ is the neighbors of node $i$ and $M$ is the edge number. More concretely, $R\rightarrow1$ stands for collective synchronization
and $R\rightarrow0$ means disordered state or local synchronization locking state; $R_{local}\rightarrow1$ shows local or collective synchronization
and $R_{local}\rightarrow0$ represents chaotic state. Thus, we distinguish the
phase locking state by synthesizing $R\rightarrow0$ and $R_{local}\rightarrow1$.

\section{Experimental results}

\subsection{Performance on artificial networks}

To test performance of the phase locking method, we construct several artificial networks with various
community structures, which are measured by modularity \cite{newman2006}
\begin{equation}\label{modularity}
Q=\frac{1}{2M}\sum\limits_{ij}{({{a}_{ij}}-{{p}_{ij}})\delta ({{C}_{i}},{{C}_{j}})},
\end{equation}
where $a_{ij}$, $p_{ij}$, $C_{i}$ and $M$ are the element of adjacent matrix $A$, $p_{ij}={{k}_{i}}{{k}_{j}}/(2M)$, community label that node $i$ belongs to and edge number, respectively.
$\delta (C_i,C_j)=1$ if $C_i=C_j$ and $\delta (C_i,C_j)=0$ otherwise.
$Q \geq 0.3$ means  significant community structure while $Q < 0.3$ represents fuzzy community structure. $Q$ in most of real networks ranges from $0.3$ to $0.7$.

The generating model is described as follows: starting from $n$ connected communities (${{U}_{1}},{{U}_{2}},....,{{U}_{n}}$) with each having some initial full connected nodes ($m_0$); at each step, a new node is added to a randomly selected community ${{U}_{l}}$ with $m$ edges; it prefers to link ${{m}_{in}}$ nodes in community ${{U}_{l}}$ and its probability to link with node $j$( $j\in {{U}_{l}}$) is $\prod{({{k}_{j}})=\frac{{{k}_{j}}}{\sum\limits_{i\in {{U}_{l}}}{{{k}_{i}}}}}$. The rest ${{m}_{out}}$ $({{m}_{out}}=m-{{m}_{in}})$ endpoints are selected from the other communities according to the similar preferential attachment described above. The artificial networks based on generating model follow power-law similar degree distribution \cite{yan2007}.

\begin{figure*}
\centering
  \subfigure[$m_{in}=6$, $m_{out}=2$, $Q=0.417$]{
    \label{network1} 
    \includegraphics[width=3in]{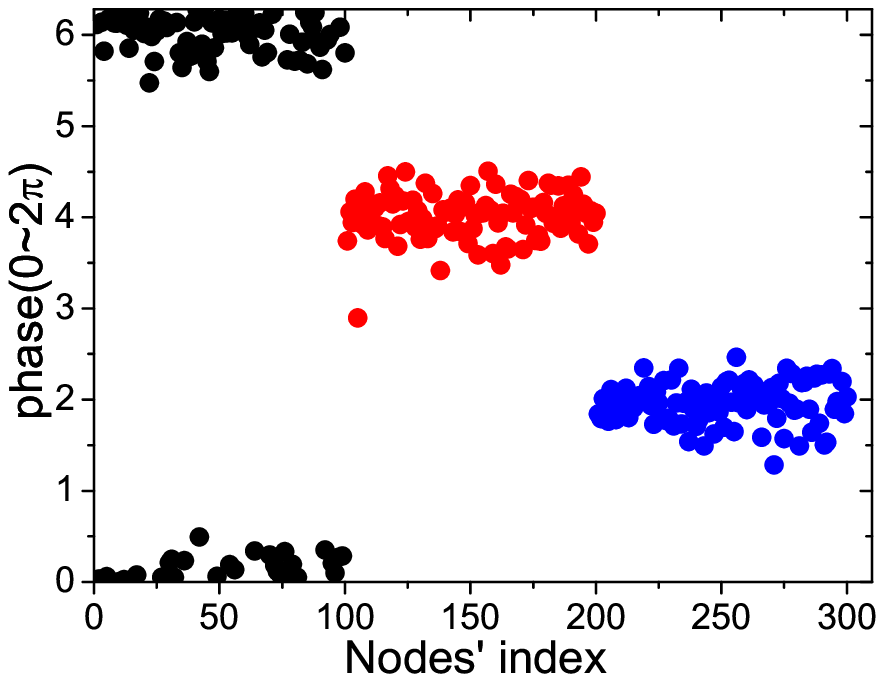}
    }
    \subfigure[$m_{in}=6$, $m_{out}=2$, $Q=0.417$]{
    \label{network1} 
    \includegraphics[width=3.1in]{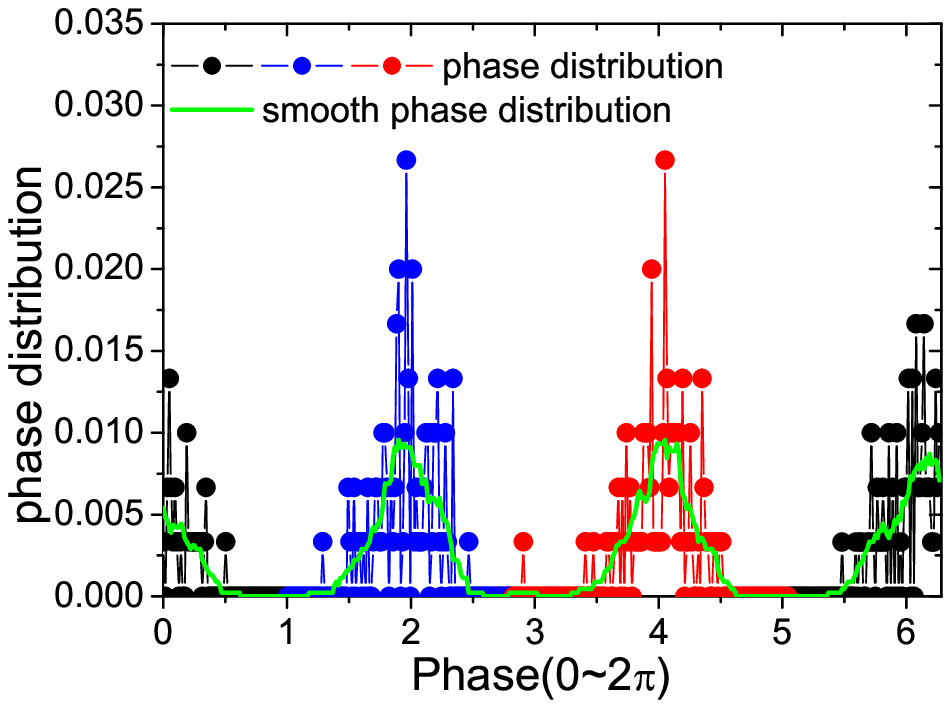}
    }
     \subfigure[$m_{in}=5$, $m_{out}=3$, $Q=0.292$]{
    \label{network1} 
    \includegraphics[width=3in]{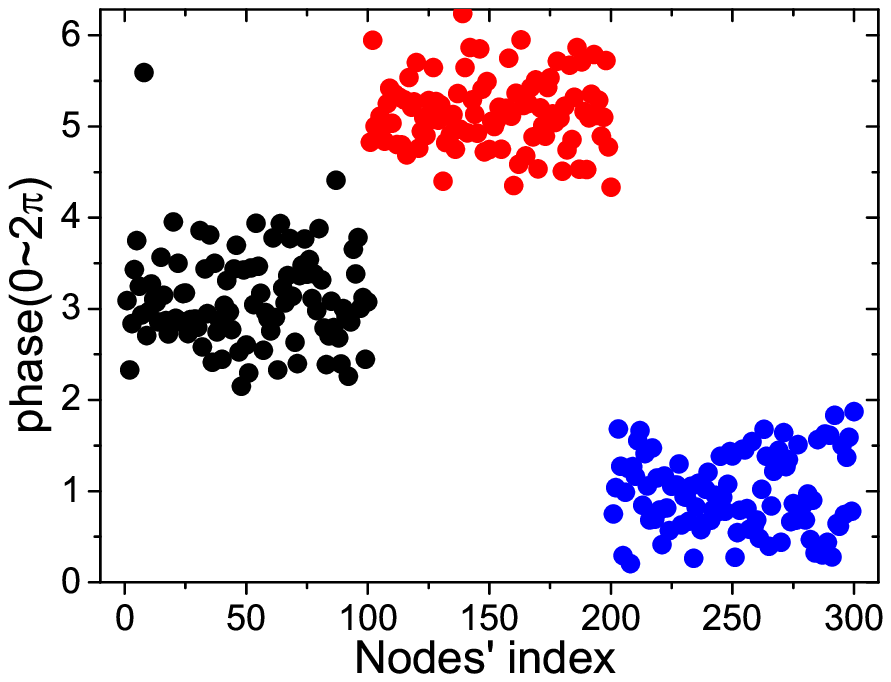}
    }
    \subfigure[$m_{in}=5$, $m_{out}=3$, $Q=0.292$]{
    \label{network1} 
    \includegraphics[width=3.1in]{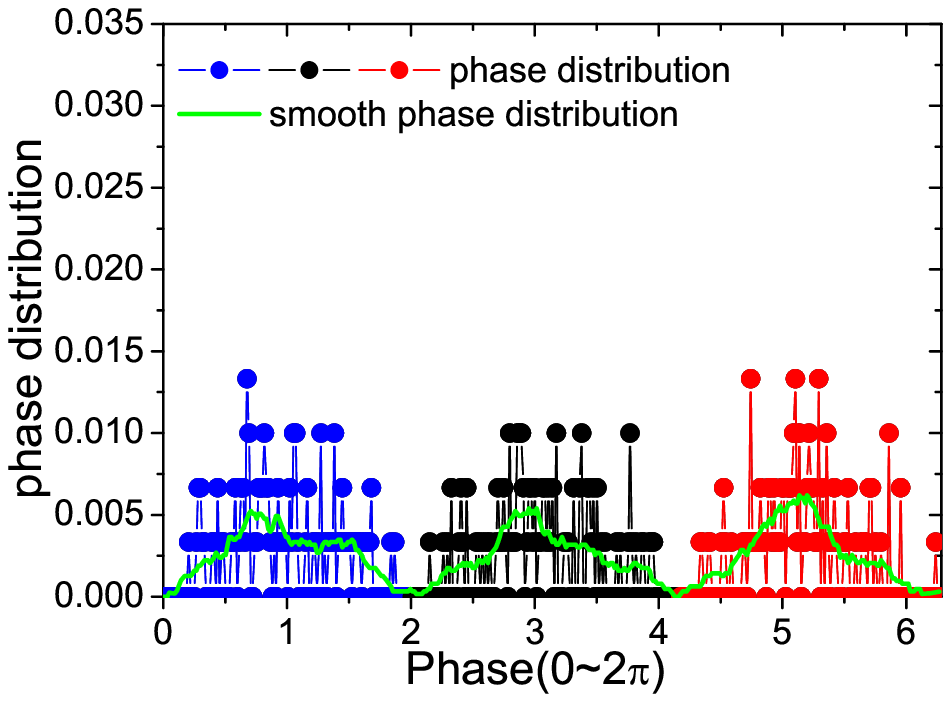}
    }
     \subfigure[$m_{in}=4$, $m_{out}=4$, $Q=0.167$]{
    \label{network1} 
    \includegraphics[width=3in]{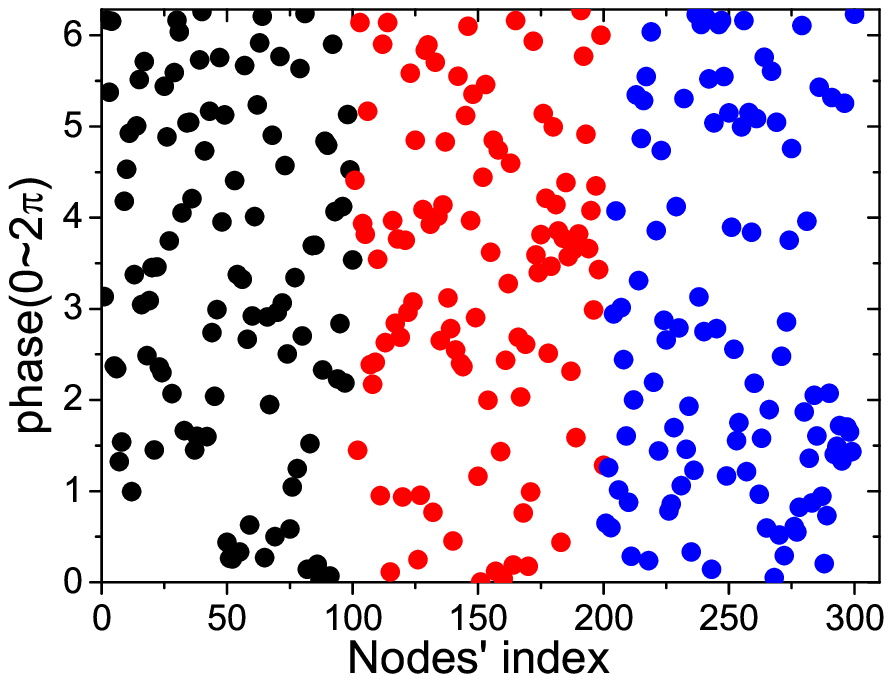}
    }
    \subfigure[$m_{in}=4$, $m_{out}=4$, $Q=0.167$]{
    \label{network1} 
    \includegraphics[width=3.1in]{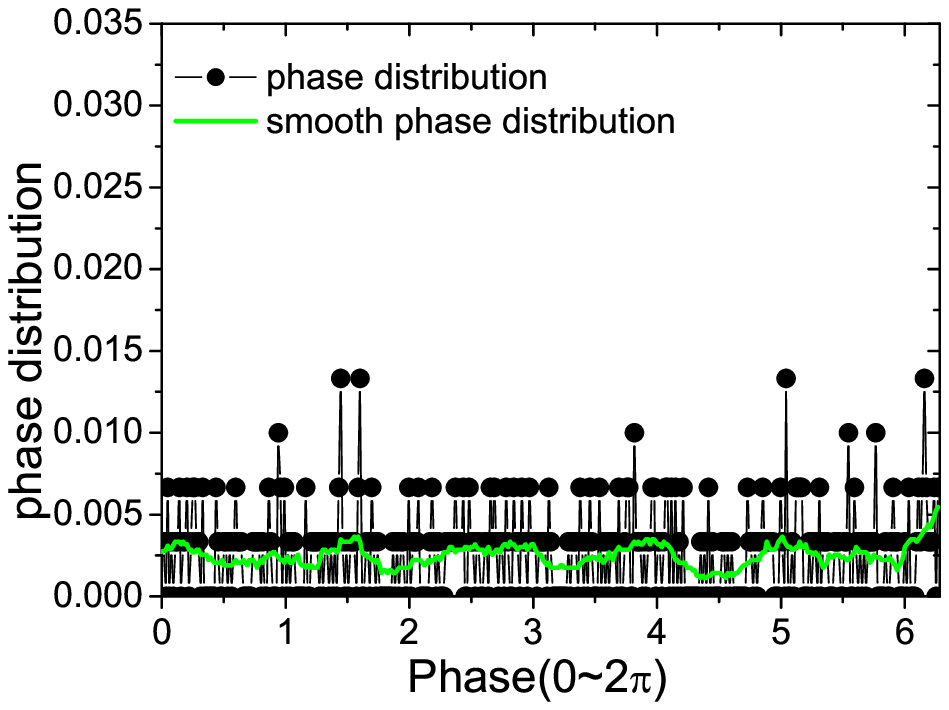}
    }
  \caption{(Color online) Nodes' final phases ($T=30$) for artificial scale-free networks with different community structures. Each new nodes links $m_{in}$ nodes in the community to which it belongs and  $m_{out}$ nodes in the other communities. The network size is N = 300. Nodes with index 1$\sim$100, 101$\sim$200
and 201$\sim$300 belong to community U1, U2 and U3 respectively. The strengths of community structures are shown in the caption of each subgraph. Subgraphs (a), (c) and (e) are the nodes'
  final phases. Subgraphs (b), (d) and(f) are corresponding phase distributions of (a), (c) and (e), respectively.
  The x-axis of subgraphs (b), (d) and(f) are all divided into 400 bins from 0 to $2\pi$ in the statistic process and the smooth lines are the average of 20 nearest neighbors.}
  \label{modelresult}
\end{figure*}

In the simulation of each artificial network, network size is $N=300$. Nodes' index $1\thicksim100$, $101\thicksim200$ and $201\thicksim300$ respectively belong to community $U_1$, $U_2$ and $U_3$, and their initial phases obey an uniform distribution ($0\thicksim 2\pi$). Figure \ref{modelresult} depicts the nodes' final phases and corresponding phase distributions for these artificial networks with various community structures. It can be found that the phases of nodes belong to different communities are clearly separated for
strong community structures (see Fig. \ref{modelresult}(a)-(d)) while the phase still follows a uniform distribution
after phase locked synchronization for very weak community structure, which proved that network with weak and no
community structure stays in disorder as the collective synchronization is suppressed
(see Fig. \ref{modelresult}(e) and (f)).
Moreover, the corresponding phase distributions fluctuate much more smoothly due to
local synchronization gradually weakening when the community structure becomes fuzzy (i.e., $Q$ decreases from 0.417 to 0.167.
As shown in Fig. \ref{modelresult}(a)-(d), the nodes at the boundaries between communities are indicated by their phases allocating at the interspace among the interval corresponding to different communities.

\begin{figure}[h]
\centering
\includegraphics[width=3in]{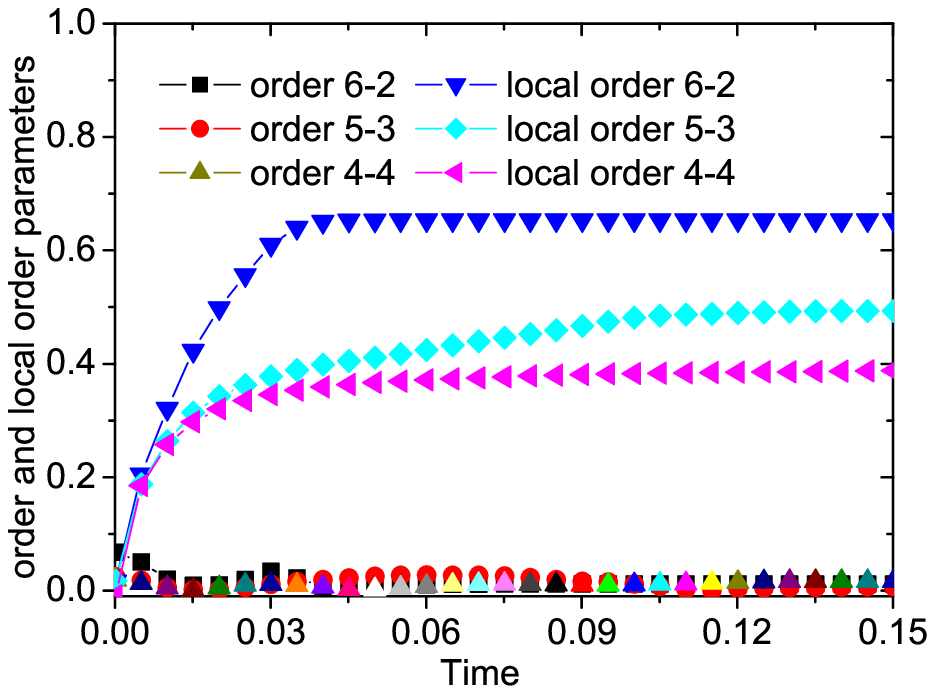}
\caption{(Color online) Evolution of the order parameter and local order parameter for different community structure. Networks are indicated as $m_{in}-m_{out}$, e.g., the symbols $6-2$ means $m_{in}=6$ and $m_{out}=2$.}
  \label{orderparameter}
\end{figure}

\begin{figure}
  \includegraphics[width=3in]{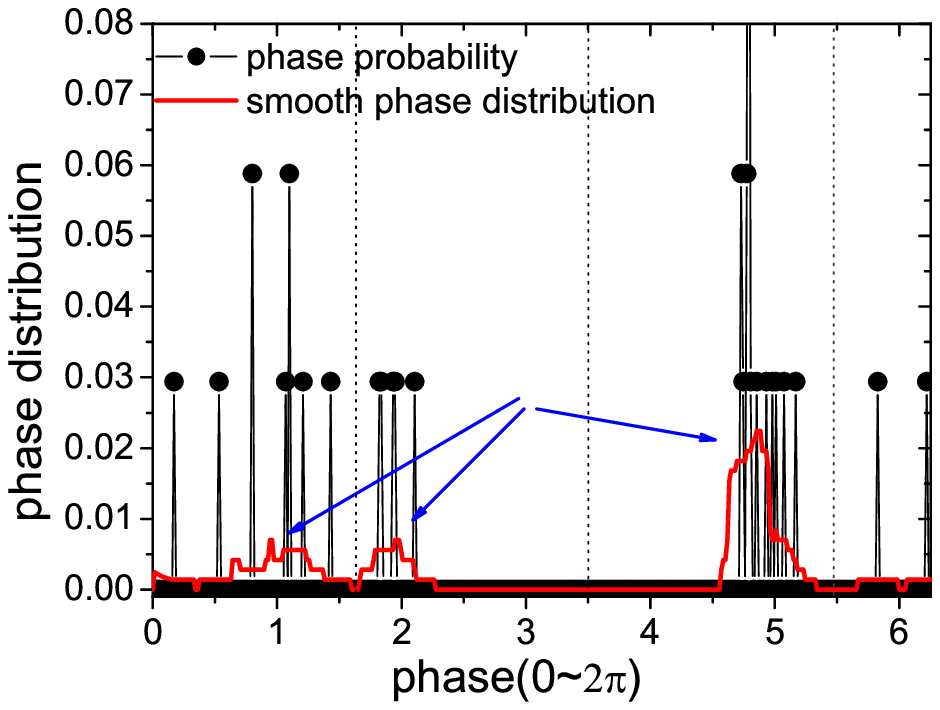}\\
  \caption{(Color online) Karate network phase distribution. Three main peaks are pointed by blue arrows and dot lines are the cut points for clustering nodes.}
  \label{karatephasedistribution}
\end{figure}

To further unveil the local synchronization paths, we present
the evolution of the order parameter $R$ and local order parameter $R_{local}$ in time scales in Fig. \ref{orderparameter}. Networks are indicated as $m_{in}-m_{out}$, e.g., the symbols $6-2$ means $m_{in}=6$ and $m_{out}=2$.
As shown in Fig. \ref{orderparameter}, $R\rightarrow0$ and
$R_{local}\gg0$
suggest that
none of the three networks can reach strong collective synchronization due to
the suppressing scheme described in Eq.\ref{mykuramoto}, but can steadily keep local synchronization.
According to the aforementioned hypothesis, the theoretic result indicating
the
networks with
strong
community structure
(the 6-2 and 5-3 networks) reaching perfect
phase locking
are $R\rightarrow0$ and $R_{local}\rightarrow1$.
However, in 4-4 network, whose community structure is weak, the $R_{local}$ is also considerably larger than $0$. This is because nodes in large amount of lines, triangles, and some other small compact groups usually synchronize and their synchronized phases increase the value of $R_{local}$.
We note that the network with more significant community structure strongly associates with a higher $R_{local}$, suggesting
stronger
local synchronization when
$Q$ increasing.
Thus,
combining results
in Figs. \ref{modelresult} and \ref{orderparameter},
we can infer the stable state of
phase locking
and detect communities
via clustering nodes' phases.

\subsection{Performance on real networks}
We have verified the efficiency of the phase locking method in
artificial networks. How does it perform on real networks? We use
five real networks from various fields, the Karate \cite{ww1977},
as-Caida, ca-GrQc, CA-HepTh, and wiki-vote network \cite{snapdatasource}
to further implement the testing experiment.
The statistical properties of these networks are illustrated in Table \ref{networkdescription}. We preprocess these network to filter some isolate nodes and many whisker ones with
degree being 1 and ignore edge direction. Then to reduce the computational
complexity we also extract the 5-core sub-networks by iteratively removing the nodes whose
degree are less than 5, except for the network of Karate.
Although we only use subnetworks of the $K-core$ parts, they can also reflect
real network property due to network self-similarity \cite{lec2009}.
\begin{table*}
\caption{\label{networkdescription}Network description of five real networks.}
\begin{ruledtabular}
\begin{tabular}{p{1in}p{2in}p{0.5in}p{0.5in}p{0.7in}p{0.5in}p{0.5in}}
 Networks&description&directed&network size&edge number&5-core&5-core edges
\\ \hline
 Karate&Karate club relation network&false&34 &78&-&- \\
 as-Caida&CAIDA AS graph from November 5 2007&true
 &26475&106762&1192&9172\\
ca-GrQc&General Relativity and Quantum Cosmology collaboration network&false&5242& 	 28980&848&6269\\
CA-HepTh&Collaboration network of Arxiv High Energy Physics Theory category&true&9877&51971&2015&10690\\
Wiki-Vote&Wikipedia vote network&true&7115&103689&3513&95028
\end{tabular}
\end{ruledtabular}
\end{table*}

\begin{table*}
\caption{\label{resultsyn}Clusters detected via phase locking($PL$) and $FASTQ$ methods.}
\begin{ruledtabular}
\begin{tabular}{p{0.65in}p{0.5in}p{0.5in}p{0.5in}p{0.5in}p{0.5in}p{0.65in}p{0.5in}p{0.5in}p{0.5in}p{0.5in}p{0.5in}}
&Network Size&$Q$($PL$)&Cluster Number ($PL$)&Largest Cluster Size($PL$)&Smallest Cluster Size($PL$)&$Q$ ($FASTQ$)&Cluster Number ($FASTQ$)&Largest Cluster Size($FASTQ$)&Smallest Cluster Size($FASTQ$)\\
\colrule
Karate&34&0.449&3&18&5&0.430&3&17&8\\
as-Caida&1192&0.284&4&652&8&0.314&8&410&2\\
ca-GrQc&848&0.787&14&149&9&0.746&25&308&5\\
CA-HepTh&2015&0.586&14&722&5&0.615&29&630&6\\
Wiki-Vote&3513&0.404&4&1483&3&0.308&3&1703&147\\
\end{tabular}
\end{ruledtabular}
\end{table*}

\begin{figure*}
\centering
  \subfigure[Cluster detected by phase locking for Karate]{
    \label{karatenetwork} 
    \includegraphics[width=3in]{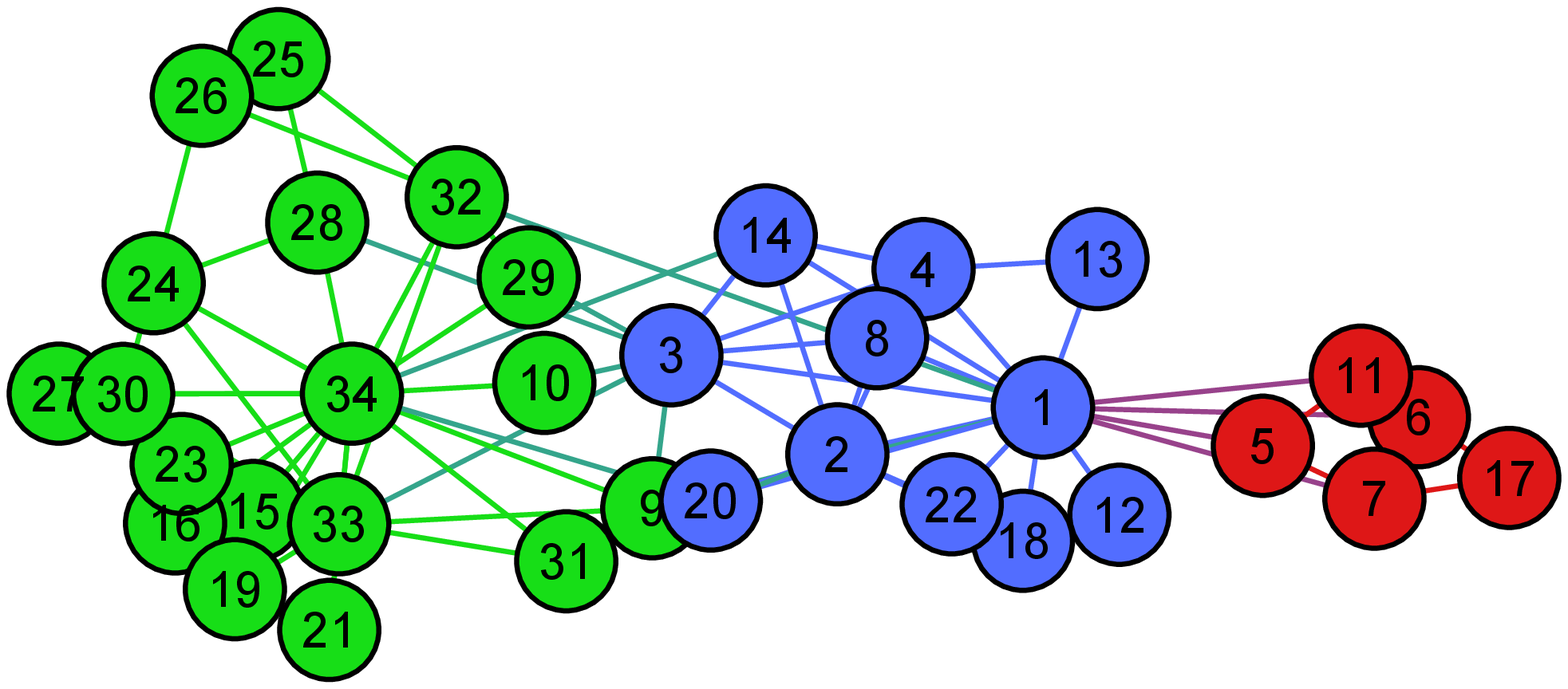}
    }
    \subfigure[Clusters detected by $FASTQ$ for Karate]{
    \label{katatenmi} 
    \includegraphics[width=3.1in]{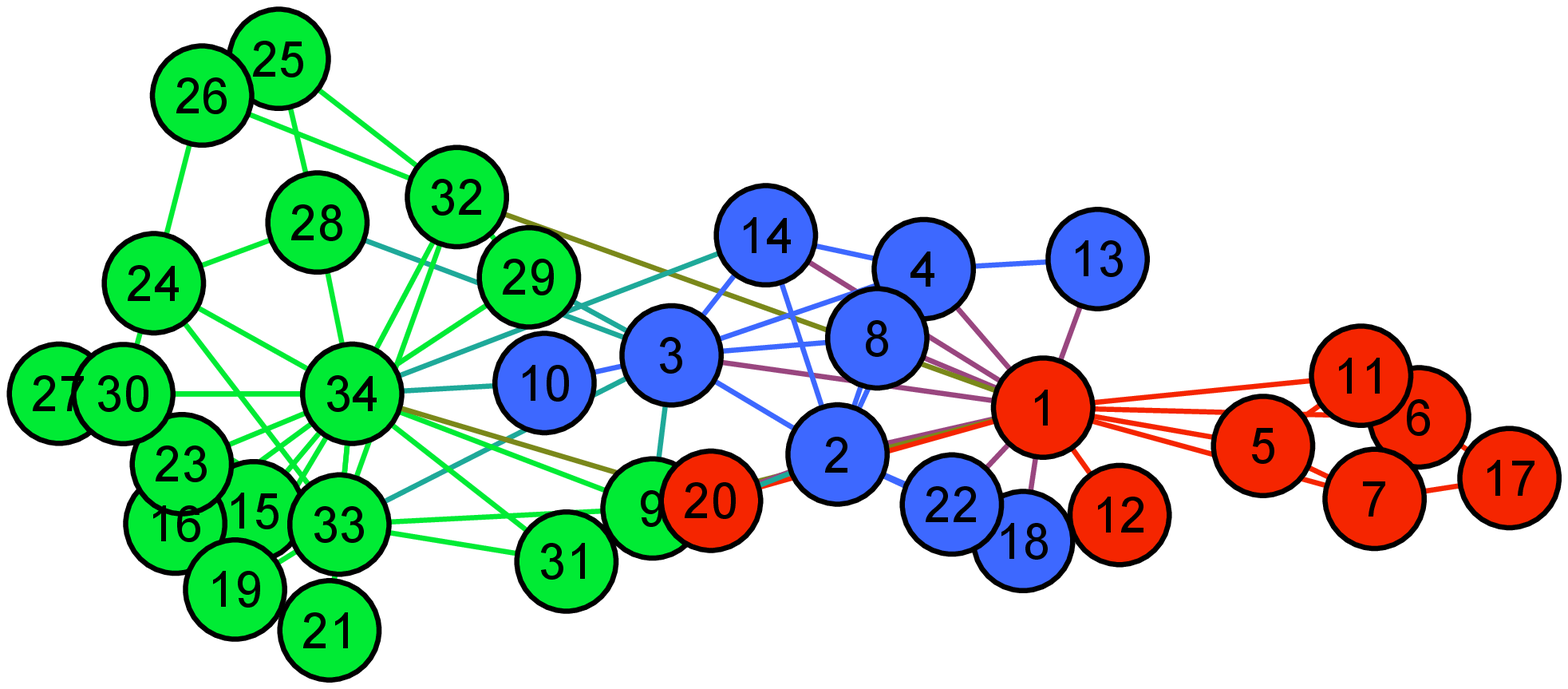}
    }
     \subfigure[Clusters detected by phase locking for ca-GrQc]{
    \label{asqrqcsyn} 
    \includegraphics[width=3in]{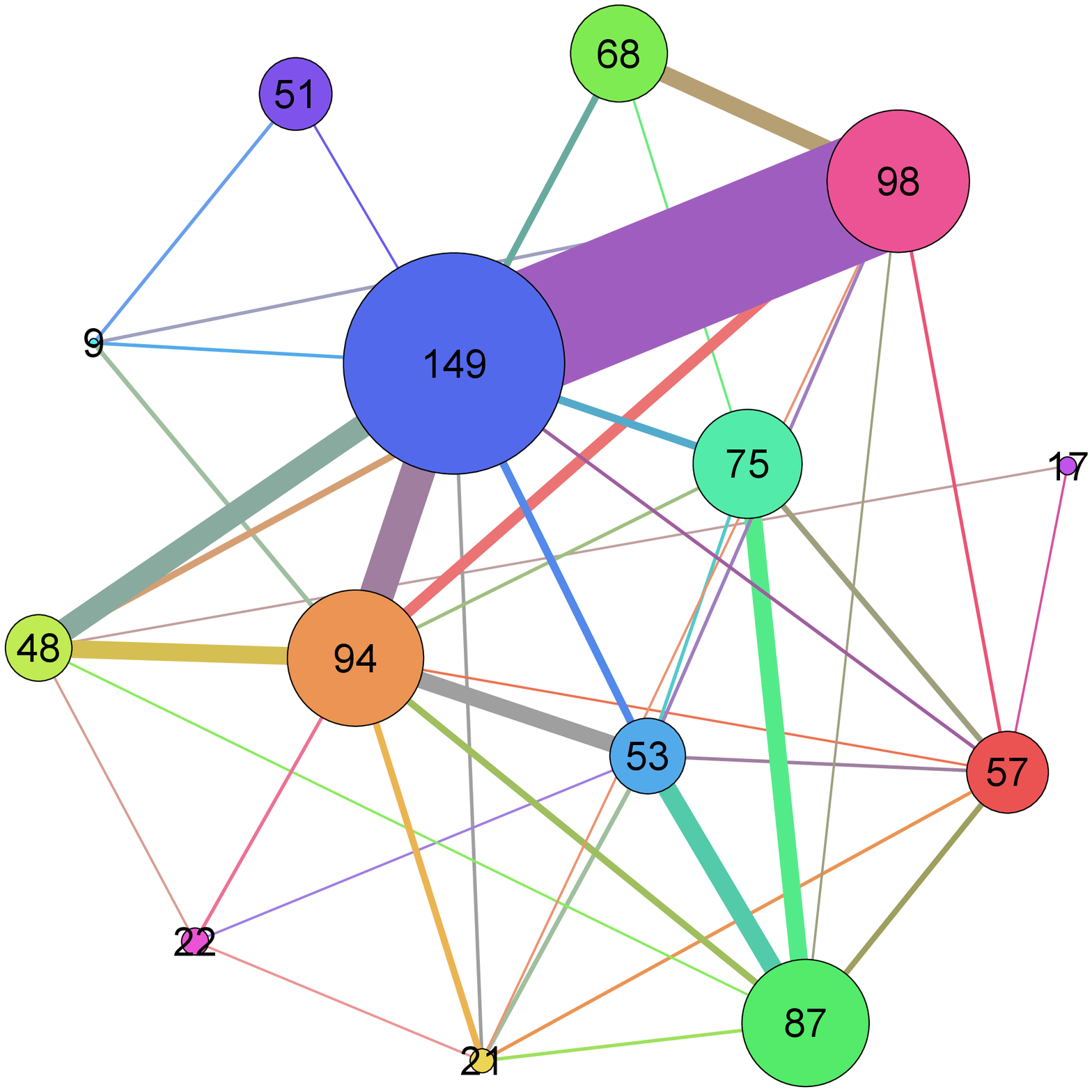}
    }
    \subfigure[Clusters detected by $FASTQ$ for ca-GrQc]{
    \label{asqrqcgn} 
    \includegraphics[width=3.1in]{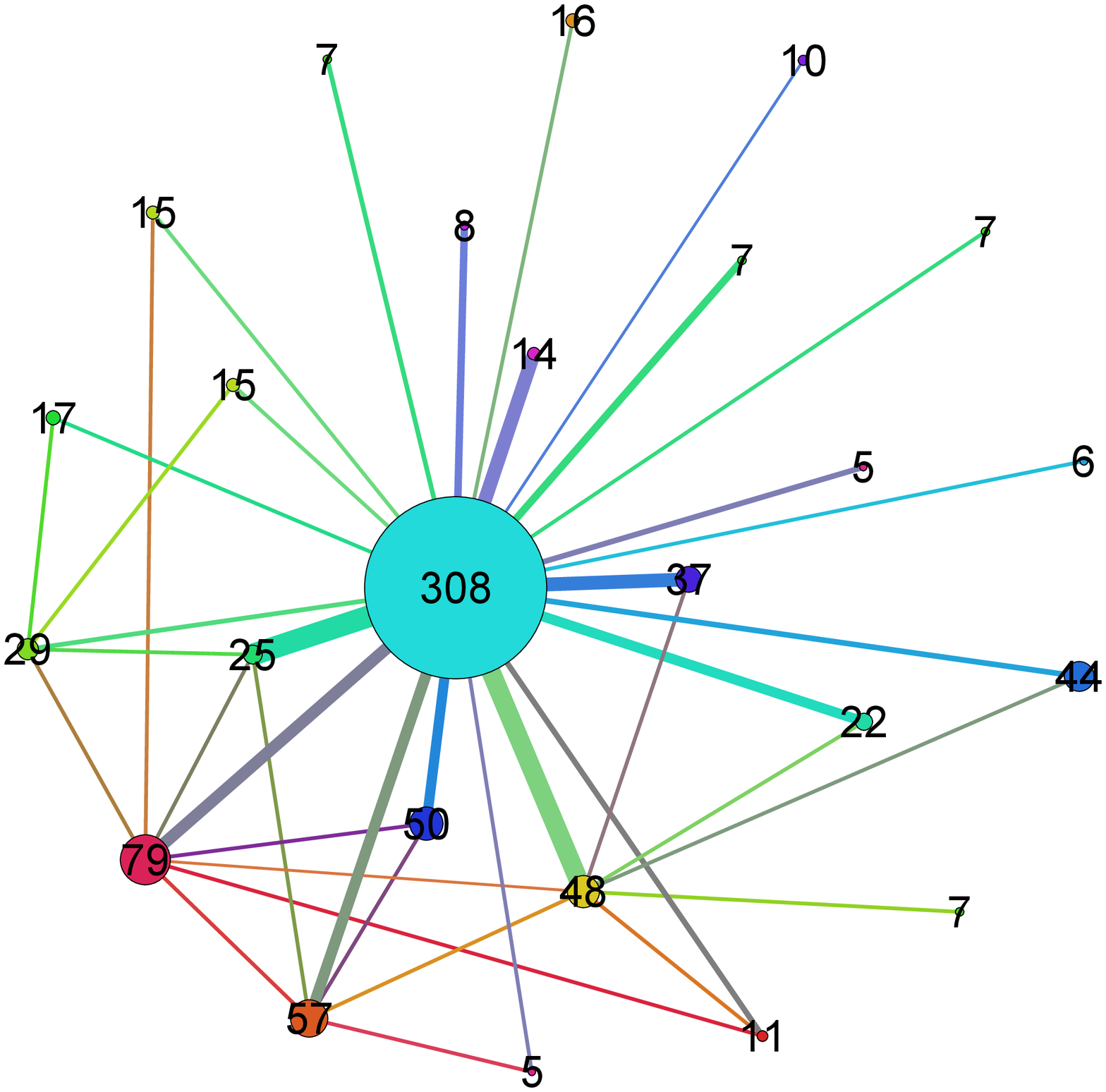}
    }
  \caption{(Color online) Community divisions of Karate and ca-GrQc networks by phase locking and $FASTQ$ methods. The numbers in (a) and (b) represent node labels; whereas number in each node represents community size in (c) and (d). Besides, node size and edge weight are proportional to the community size and relation between communities in (c) and (d), respectively. }
  \label{karatelocalsyn}
\end{figure*}

After enough synchronizing time ($T=30$), synchronization reaches stable state, nodes in the same clusters synchronize together and form peaks in the phase distributions. We roughly cluster phases at the valleys of smoothed phase distribution waveform (e.g., Karate network shown in Fig. \ref{karatephasedistribution}). Influenced by the community size, the smoothed phase distribution waveform has
much fluctuation noise and some nodes on the community boundary may be improperly classified.
Thus, the partitions are regulated to ensure that every node has more links inner the community
to which it belongs than any other communities.
Table \ref{resultsyn} shows the results of detecting clusters
through the phase locking and referential modularity-based fast algorithm which is a greedy agglomerating method to
detect communities by maximizing the modularity at each step, proposed by Newman and his colleagues ($FASTQ$ for short) \cite{newman2006, newman2004}. We choose $FASTQ$ as a reference for its simplicity and high speed. Synthesizing the
results of phase locking and $FASTQ$ in Table \ref{resultsyn}, we can observe that the values of
modularity $Q$ obtained from phase locking and $FASTQ$ are close on all five real networks, but there
are also differences in cluster number, the largest and smallest cluster size. In
particular, the modularity isn't the only criterion to estimate community detection algorithms. Among
those criterions, the most important is that clusters must be in consistency with reality. In real world,
approximate cluster size ranges from dozens to hundreds \cite{lec2009}, thus a cluster whose size is less
than 10 is too small to be a reasonable community. Moreover, recent study \cite{kehagias2012} shows
that the best reasonable division for community detection may not fix with the highest $Q$, and for
these modularity optimization methods large clusters can easily swallow small ones due to the definition
of modularity. Herein, we need to look into the detail of community sizes to distinguish the quality of
different community divisions by phase locking and $FASTQ$ for the same network.


\begin{figure*}
\centering
  \subfigure[as-Caida]{
    \label{networksize1} 
    \includegraphics[width=3in]{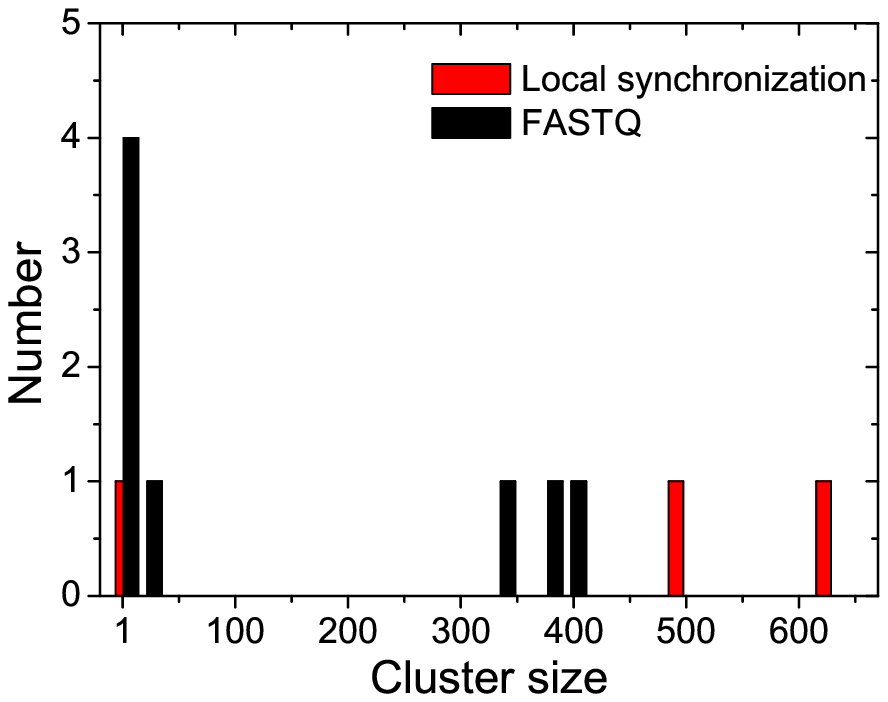}
    }
    \subfigure[ ca-GrQc]{
    \label{networksize2} 
    \includegraphics[width=3.1in]{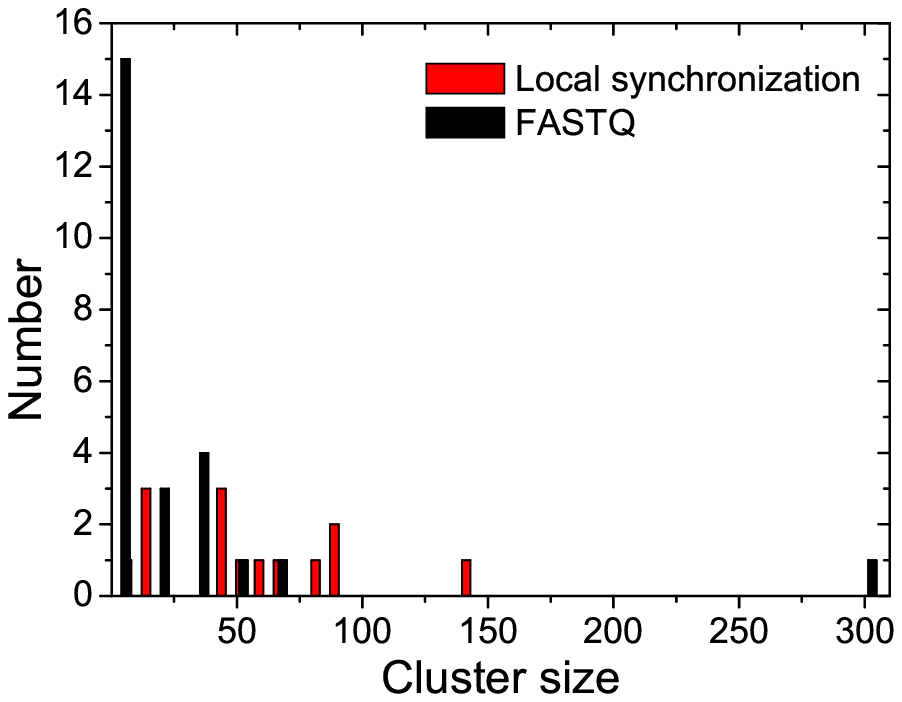}
    }
     \subfigure[ CA-HepTh]{
    \label{networksize3} 
    \includegraphics[width=3in]{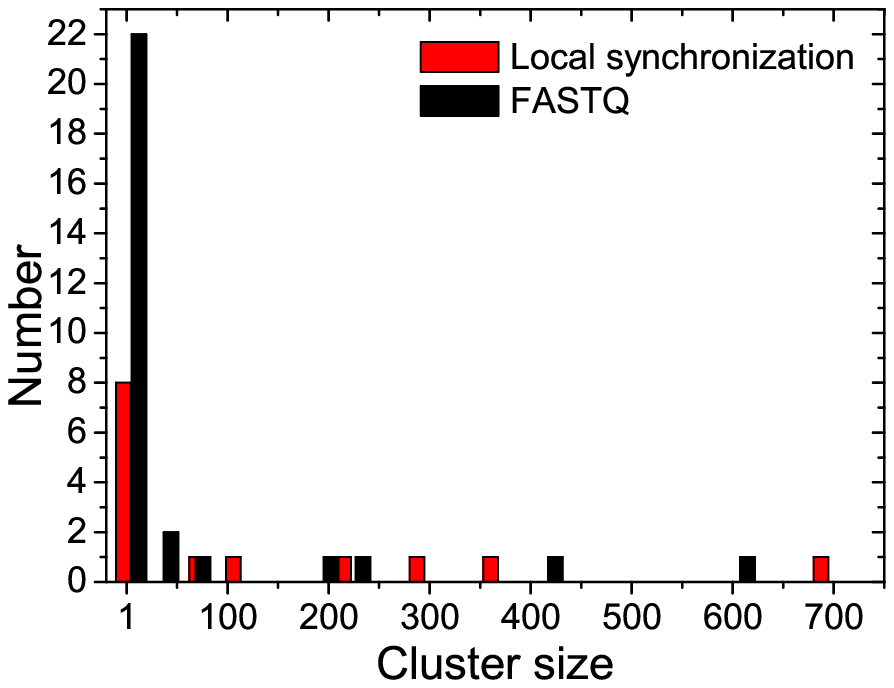}
    }
    \subfigure[Wiki-Vote]{
    \label{networksize4} 
    \includegraphics[width=3.1in]{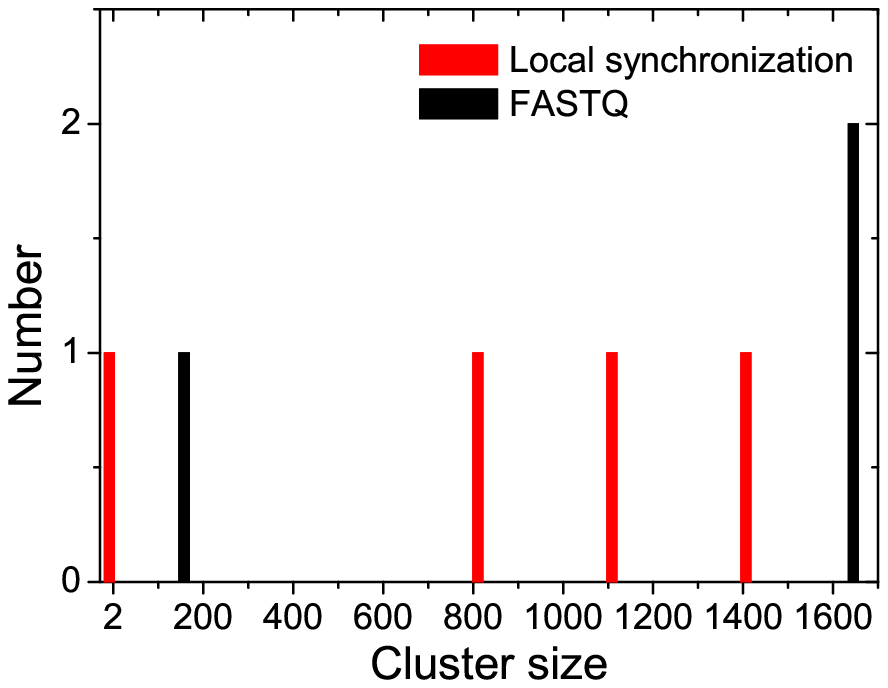}
    }
  \caption{(Color online) Cluster size distributions of as-Caida, ca-GrQc, CA-HepTh and Wiki-Vote. Red and black columns are results of phase locking an $FASTQ$ methods, respectively. The x-axis is divided into 20 bins from 1 to maximum community size and the amplitude of histogram is proportional to the number of communities whose size locates in the bin. }
  \label{clustersize}
\end{figure*}

More concretely, taking Karate and ca-GrQc networks as examples, their dividing results are visualized explicitly
in Fig. \ref{karatelocalsyn}. First, as shown in Fig. \ref{karatenetwork}, groups of nodes
in different colors represent three communities detected by phase locking, which is in
consistency with the three peaks in Fig. \ref{karatephasedistribution}.
In particular, the green nodes form a larger and more compact cluster that results in
the highest peak ($phase\approx 4.8$). Though $FASTQ$ method also divides Karate network into three communities
(shown in Fig. \ref{katatenmi}), there are some distinct differences of cluster divisions
from phase locking to $FASTQ$ methods. Note that, nodes labeled $1$, $10$, $12$, $20$ are classified into
different communities for the two algorithms. In real world, Karate network represents the
friendships in a university karate club, which split into two groups
because the conflict about the club fees. In the club, node $1$ indicates the karate instructor who wanted
to raise the fees and node $33$ indicates the club's chief administrator wanted to stabilize the fees.
They both insisted their own opinions and became the leaders of the two separate groups, which makes
nodes $1$ and $33$ the community cores of the \emph{Karate} network.
Several people (nodes on the border of two communities)
were not opposing either side
and kept friendship with both groups. Thus, both organization wanted to draw over these
people to stand their own position and the phenomenon can also be observed in Fig. \ref{karatephasedistribution},
in which the phases of those uncertain nodes locate at the interspace among their neighbors. From Fig. \ref{karatenetwork}, it's clear that phase locking method distinguishes the two clusters and all nodes
on the border are partitioned right. Besides the two communities, the red nodes (Node $5$, $6$, $7$, $11$
and $17$) also forms a cluster. Those nodes are more compact than the blue clusters and split out to
form a new community which
disagrees with the reality but satisfies the definition of community structure.
While $FASTQ$ method partitions the network imprecisely and some nodes are
improperly
classified with lower $Q$ in Fig. \ref{katatenmi}. Node $10$ and $1$ are categorized
in wrong communities: people represented by nodes $10$ tended to support node $33$, which leads to that
$10$ should belong to the group leaded by node $33$;
people labeled $1$ was the
core of one group and should belong to the same group with node $8$ and $13$.
Thus, phase locking behaves better than $FASTQ$ in Karate club network. What's more, figures. \ref{asqrqcsyn} and \ref{asqrqcgn} are the division results of phase locking and $FASTQ$ methods for ca-GrQc network, in which
each node stands for a cluster and node size is proportional to the community size. Clusters detected
by phase locking have more balanced cluster size ranging from dozens to
hundreds(See in Figs. \ref{asqrqcsyn} and \ref{asqrqcgn}).

Figure \ref{clustersize} shows the cluster size distribution for both algorithms vividly.
In Fig. \ref{clustersize},
plenty of communities detected by $FASTQ$ have small cluster sizes
less than $10$ that deviate the nature community size. Whereas, phase locking
method divides networks into more balanced communities than $FASTQ$ for the
four networks. Combining results shown in
Table \ref{resultsyn} and Fig. \ref{clustersize} we can draw the conclusion
that though the two algorithms scores similar modularity $Q$, phase locking method divides more
resealable and even communities and the detected communities corresponds to nature better compared
with those of $FASTQ$.

\section{Conclusion}
In conclusion,
we assume that the phase synchronization is comprised of local
and collective dynamics and infer
that the networks with inapparent community structure
(or random network) tend to synchronize through collective dynamics while
the networks with significant community structure synchronize through both local and collective
dynamics. Based on this idea, we propose a scheme to suppress collective dynamics of network
via the comparison of original network and its corresponding first-order null model network
(see in Eq. \ref{mykuramoto}) and lock nodes' phases into stably local dynamics of network.
Through this scheme, the community structure of network is able to be significantly unveiled
when nodes' phases are locked. Firstly, the
efficiency of the scheme is proved by the experimental results on artificial network. That is, the order
parameter $R$ is close to $0$ and the local order parameter $R_{local}$ is strongly positive
correlation with modularity $Q$ and converge to $1$, meanwhile the nodes in
the same community synchronize together and phases of nodes in different communities are clearly
separated. Secondly, we apply this scheme to five real networks and obverse
a better division of network in comparison with $FASTQ$ (e.g.,
visualization of network in Fig. \ref{karatelocalsyn}) and more reasonable cluster size distribution (see in
Fig. \ref{clustersize}). Finally, It is worth to be mentioned that the overlapping nodes among
communities are usually located at the valleys of phase distribution
curve, thus the overlapping communities are able to be distinguished based on this scheme. Besides,
the novel synchronization method can be modified to accommodate networks with hierarchical
structure and reveal community structure at different hierarchical levels, which is similar to
the mesoscopic analysis of network topology \cite{Gomez2008,Granell2011}, which will be the future works.

\section{Acknowledgement}
This work is jointly supported by the National Nature Science Foundation of China (Nos. 60974079 and 61004102),
China Postdoctoral Science Foundation (No. BH2100100014), and the Fundamental Research Funds for the Central Universities
(No. ZYGX2012J075)


\begin{thebibliography}{00}

\bibitem{santofortunato2010} S. Fortunato, Phys. Rep. 486, 75 (2010).
\bibitem{porter2009} M. A. Porter, J. P. Onnela, and P. J. Mucha, Not. Am. Math. Soc. 56, 1082 (2009).
\bibitem{cocia2011} M. Coscia, F. Giannotti, and D. Pedreschi, Stat. Anal. Data Mining 4, 512 (2011).
\bibitem{honglei2012} L. Hong, S. M. Cai, J. Zhang, Z. Zhuo, Z. Q. Fu, and P. L. Zhou, Chaos 22, 033128 (2012).
\bibitem{wangbiao2013}B. Wang, Z. Zhuo, S. M. Cai, and Z. Q. Fu, Physica A 392, 1902 (2013).
\bibitem{zhoum2012}M. Y. Zhou, S. M. Cai, and Z. Q. Fu, Physica A 391, 1887 (2012).
\bibitem{newman2004} M. E. J. Newman, Phys. Rev. E 69, 066133 (2004).
\bibitem{palla2005} G. Palla, I. Der\'{e}nyi, I. Farkas, and T. Vicsek, Nature 435, 814 (2005).
\bibitem{ahn2010}Y. Y. Ahn, J. P. Bagrow, and S. Lehmann, Nature 466, 761 (2010).
\bibitem{newman2006} M. E. J. Newman, Proc. Natl. Acad. Sci. U.S.A. 103, 8577 (2006).
\bibitem{lec2009} J. Leskovec, K. J. Lang, A. Dasgupta, and M. W. Mahoney, Internet Math. 6, 29 (2009).
\bibitem{good2010} B. H. Good, Y. A. de Montjoye, and A. Clauset, Phys. Rev. E 81, 046106 (2010).
\bibitem{kehagias2012} A. Kehagias, arXiv:1209.2678 (2012).
\bibitem{albert2012} A. L. Barab\'{a}si, Nature 489, 507 (2012).
\bibitem{arenas2008} A. Arenas, A. D¨ªaz-Guilera, J. Kurths, Y. Moreno, and C. Zhou, Phys. Rep. 469, 93 (2008).
\bibitem{wu2012} J. Wu, L. Jiao, C. Jin, F. Liu, M. Gong, R. Shang, and W. Chen, Phys. Rev. E 85, 016115 (2012).
\bibitem{kim2010} Y. Kim, Y. Ko, and S. H. Yook, Phys. Rev. E 81, 011139 (2010).
\bibitem{arenas2006} A. Arenas, A. D\'{\i}az-Guilera, and C. J. P\'{e}rez-Vicente, Phys. Rev. Lett. 96, 114102 (2006).
\bibitem{toni2011} T. P\'{e}rez, V. M. Egu\'{\i}luz, and A. Arenas, Chaos 21, 025111 (2011).
\bibitem{jesus2007} J. G\'{o}mez-Gardenes, Y. Moreno, and A. Arenas, Phys. Rev. Lett. 98, 034101 (2007).
\bibitem{bohm2010} C. B\"{o}hm, C. Plant, J. Shao, and Q. Yang, Proceedings of the 16th ACM SIGKDD international conference on knowledge discovery and data mining, ACM, p.583, 2010.
\bibitem{Gomez2008} A. Arenas, A. Fernandez, and S. Gomez, New. J. Phys. 10, 053039 (2008).
\bibitem{Granell2011} C. Granell, S Gomez and A. Arenas, Chaos 21, 016102 (2011).
\bibitem{Granelltools} \url{http://deim.urv.cat/~sgomez/radatools.php} Toolbox for community detection.
\bibitem{yan2007} G. Yan, G. Chen, J. Lu, and Z. Q. Fu, Phys. Rev. E 80, 1 (2009).
\bibitem{ww1977} W. W. Zachary, J. Anthropol. Res. 33, 452 (1977).
\bibitem{tong2008} H. Tong, C. Faloutsos, and J. Y. Pan, Knowl. Inf. Syst. 14, 327 (2008).
\bibitem{snapdatasource} \url{http://snap.stanford.edu/data/index.html} Source of Datasets.
\end{thebibliography}

\end{document}